\documentstyle[preprint,aps]{revtex}                  
\begin{document}
\draft

\preprint{MKPH-T-95-21}
\title{Elastic Pion Scattering on the Deuteron
in a Multiple Scattering Model}
\author{S. S. Kamalov\cite{Sabit} and L. Tiator}
\address{Institut f\"ur Kernphysik, Universit\"at Mainz, 55099 Mainz, Germany}
\author{C. Bennhold}
\address{Center for Nuclear Studies, Department of Physics, The George
Washington University, \\Washington, D.C., 20052, USA}

\maketitle

\begin{abstract}
Pion elastic scattering on deuterium is studied in the
KMT multiple scattering approach developed in momentum space. 
Using a Paris wave function and the same methods and approximations as commonly
used in pion scattering on heavier nuclei 
excellent agreement with differential cross section data is obtained for a 
wide range of pion energies. Only for $T_{\pi}>250$ MeV and very backward 
angles, discrepancies appear that are reminiscent of disagreements in pion 
scattering on $^3$He, $^3$H, and $^4$He.
At low energies the second order corrections have been included.
Polarization observables are studied in detail. While tensor analyzing powers
are well reproduced, vector analyzing powers exhibit dramatic
discrepancies.
\end{abstract}


\section{INTRODUCTION} 

In the realm of pion-nuclear physics,
the pion interaction with the deuteron is of special
interest\cite{Garc}. 
On the theoretical side, it is the simplest few-body system 
in which conventional theories (Faddeev theory, coupled $\pi NN - NN$ 
equations and multiple scattering theory) can be rigorously tested. 
The knowledge obtained from these studies can then be extended to
many-body systems. On the experimental side, a large set 
of measurements for differential cross sections and polarization 
observables are available in $\pi d$ scattering. Thus, there is sufficient
motivation to study pion interaction mechanisms with very light nuclei in 
detail.
 
In recent years the three-body approach has been used extensively
in describing the pion-deuteron  interaction. From the point of view of 
solving the $\pi NN$ three-body problem directly this is clearly 
an important achievement. However, due to the complexity of the three-body 
formalism 
the application of the obtained results to the case of pion interaction with 
heavier nuclei may be more difficult.

On the other hand, a microscopic description of the pion-nuclear interaction
in the framework of the multiple scattering theory \cite{KMT,Watson}
has continuously been improved. More recent developments are
due especially to techniques developed to perform calculations
in momentum space
using the KMT\cite{KMT} formulation of the multiple scattering theory. 
In momentum space
nonlocalities of the pion-nuclear interaction, off-shell extrapolations of 
the pion-nucleon scattering amplitudes and exact treatment of Fermi motion 
can be taken into account \cite{Landau,Mach,KTB93,KTB92}. 
The KMT approach is important to avoid double counting of pion rescattering 
on the same nucleon.
With the addition of the phenomenological $\rho^2$ term~\cite{GKM},
which is responsible for real pion absorption and second 
order effects, the momentum space formalism was successful in 
describing the pion-nuclear interaction in the $\Delta$ resonance region 
as well as at low energies for a large set of nuclei with A=4--40. 
In Refs.\cite{Ciep}, this method has been extended successfully to the
description of pionic atoms as well.

The aim of the present work is a systematic investigation of pion
scattering on the deuteron in the region of 
pion kinetic energies of $T_{\pi}$=60--300 MeV.  We employ a
multiple scattering framework in momentum space using only elementary 
amplitudes extracted from $\pi N$ scattering data and realistic deuteron 
wave functions. Due to the latter ingredient we are in a good position 
to fix the nuclear structure input in order to shed more light on the 
reaction mechanism. Another purpose is to develop a pion deuteron
interaction that can be used for pion photoproduction on the deuteron in
a straightforward way. Such an approach, applied to pion scattering and 
pion photoproduction on the trinucleon~\cite{KTB92,KTB93} has provided 
an excellent description of the differential cross section. 
This has motivated us 
to extend our momentum space approach for these reactions to the deuteron. 

Our study has been divided into two parts. In this paper we investigate pion 
elastic scattering on the deuteron, testing our 
multiple scattering approach by comparing with all available experimental data. 
Coherent $\pi^0$ photoproduction is considered in the second part. 
The main aspects of our formalism based on the KMT multiple scattering
approach and coupled-channels method are given in section 2.
Section 3 presents our results for elastic pion-deuteron scattering
while our conclusions are summarized in section 4.
 
\section{FORMALISM}
 
\subsection{General overview}
 
In multiple scattering theory the pion-nuclear T-matrix
can be presented as
\begin{equation}
 T=\sum_{i=1}^A T_i(E)\,,
\end{equation}
where
\begin{equation}
T_i(E)=\tau_i(E)+\tau_i(E)G_0(E)\sum_{j \neq i}^A T_j(E)
\end{equation}
and $\tau_i(E)$ is defined as a solution of the
equation
\begin{equation}
\tau(E) = v + v G_0(E) \tau(E)\,.
\end{equation}
In Eqs. (2-3), $G_0(E)$ is the Green's function of the non-interacting
pion-nuclear system and $v$ is the potential of the pion-nucleon interaction.
Note that the $\tau$-matrix in Eq. (3) differs from the standard two-body
t-matrix for free pion-nucleon scattering through the
Green's function $G_0(E)$ which contains the many-body nuclear Hamiltonian.
For the free pion-nucleon t-matrix we have the equation
\begin{equation}
t(\omega) = v + v g_0(\omega)t(\omega) 
\end{equation}
with the free pion-nucleon Green's function $g_0(\omega)$.

Following the KMT-version of multiple scattering
theory~\cite{KMT}, Eqs. (1-4) are equivalent to the system of integral 
equations for the auxiliary matrix $T'=\frac{(A-1)}{A}T$
\begin{mathletters}
\begin{equation}
T'(E)=U'(E)+U'(E)G_0(E) P T'(E)\,,
\end{equation}
\begin{equation}
U'(E)=(A-1)\tau(E)+(A-1)\tau(E)G_0(E) Q U(E)\,,
\end{equation}
\begin{equation}
\tau(E)=t(\omega)+t(\omega)[G_0(E)-g_0(\omega)]( P+Q )\tau(E)\,,
\end{equation}
\end{mathletters}
where the operators $P=\mid 0><0\mid$ and $ Q =\sum_{m\neq 0}
\mid m><m\mid$ project the nuclear state vectors into the ground state
$\mid 0>$ and all possible excited states $\mid m>$, respectively. In the 
case of the deuteron, the operator $Q$ projects onto the states of the
continuum.
 
One of the main problems which arise in applying these equations
to pion deuteron scattering is the correct description of the
coupling to the break-up or continuum channels. In this case
the Lippmann-Schwinger equation has a noncompact kernel and,
therefore, cannot be solved numerically. It is well known that the 
solution to this problem was found by Faddeev who from the system of 
Eqs. (5) derived a new set of equations which do not have this inherent 
shortcoming.

In this paper we demonstrate that multiple scattering theory is able 
mainly to reproduce existing experimental data
using several standard approximations for the solution of the Eqs. (5).
This in turn generates a simple pion-deuteron interaction that will be
used in our description of coherent $\pi^0$ photoproduction, discussed
in the following part. As a first step, we {\em neglect the contribution 
from the coupling to the break-up channel}. This means that in Eqs. (5b-5c) 
the contribution from the
$Q$-operator is dropped. The motivation for this approximation is that
contributions from the diagonal matrix elements dominate due to the large 
overlap of the nuclear wave functions in the initial and final states.
As we will see below this approximation is a resonable starting point for
the study of pion-nuclear interaction. However, at low energies where for
isoscalar nuclei the contribution from the $ P$ operator
is very small, the coupling to the break-up channels becomes important.

The next approximation is the special choice for connecting the
pion-nucleon energy, $\omega$, with the pion-nuclear one, $E$.
In general, the energy of the pion-nucleon system is a dynamical variable 
in the many-body system~\cite{Fadd}. Thus, there are many possibilities
for relation $\omega$ and $E$; a detailed investigation of this
issue was done in Ref.\cite{Mach}. Here we follow the
prescription of Landau and Thomas~\cite{Landau}
which is called {\em modified impulse approximation}
with the three-body choice for the reaction energy. The relativistic 
generalization of this approximation in the case of the deuteron leads to the 
following expression for the pion-nucleon energy~\cite{Mach}
\begin{equation}
\omega  =  E+m_\pi+M_N- \left[ (m_\pi+M_N)^2+
\frac{(\vec{q}\,'+\vec{q})^2}{16} \right]^{\frac{1}{2}}-
\left[M_N^2+\frac{(\vec{q}\,'+\vec{q})^2}{16}\right]^{\frac{1}{2}}\,,
\end{equation}
where $m_\pi$ and $M_N$ are the pion and nucleon masses, respectively.
As shown in Refs.\cite{Landau,Gurv},
this choice minimizes the contribution from the second term in Eq. (5c).

The scattering amplitude $F_{M_f M_i}(\vec{q}\,',\vec{q})$ is connected 
with the $T'$-matrix in Eq. (5a) by the relation
\begin{equation}
 F_{M_f M_i}(\vec{q}\,',\vec{q})=
-\frac{\sqrt{{\cal M}(q'){\cal M}(q)}}{2\pi}\frac{A}{A-1}
<\pi (\vec{q}\,'),f \mid T'(E)\mid i,\pi (\vec{q})>\,,
\end{equation}
where $\vec{q}$ and $\vec{q}\,'$ are the pion momenta in the
intial and final states, respectively, $\mid i>=\mid 1^+ M_i>$ and 
$\mid f>=\mid 1^+ M_f>$ denote the nuclear initial and final states, 
with the nuclear spin projection $M_i$ in the initial and $M_f$ final 
states. The pion-nuclear reduced mass is given by 
${\cal M}(q)=E_{\pi}(q)E_{A}(q)/E(q)$, where $E(q)=E_{\pi}(q)+E_{A}(q)$ 
is the total pion-nuclear energy. 
To shorten the notation in Eq. (7) we skipped the dependence on the energy
$E$ for the scattering amplitude. For a given pion energy, $E$ is fixed and 
in the following we use $E=E(q)$.
In the same way 
we can express the pion-nuclear potential in momentum space 
$V_{M_f M_i}(\vec{q}\,',\vec{q})$ via matrix $U'(E)$ from Eq. (5b)
Then, in accordance with Eq. (5a) the elastic scattering amplitude 
can be constructed by solving the integral equation with relativistic 
kinematics
\begin{equation}
{\rm F}_{M_f M_i}(\vec{q}\,',\vec{q}) ={\rm V}_{M_f M_i}
(\vec{q}\,',\vec{q})-\frac{a}{(2\pi)^2} \sum_{M} \int \frac{d\vec{q}\,''}
{{\cal{M}}(q'')}\frac{{\rm V}_{M_f M}(\vec{q}\,',\vec{q}\,'')
{\rm F}_{M M_i}(\vec{q}\,'',\vec{q})}{E(q) - E(q'') + i\epsilon} \, \, ,
\end{equation}
where the factor $a=(A-1)/A$ is important to avoid double counting of pion
rescattering on the same nucleon, which is already included in the
free pion-nucleon $t$-matrix. Note that due to this factor, Eq. (8) 
{\em is not} the standard Lippmann-Schwinger equation.

In accordance with the approximations discussed above and after neglecting
the contributions from the second terms in Eqs. (5b) and (5c) 
the momentum space potential of the pion-nuclear interaction can be given by
\begin{equation}
 V_{M_f M_i}(\vec{q}\,',\vec{q})=
 V_{{\rm {Coul.}}}(\vec{q}\,'-\vec{q};R)\delta_{M_f M_i}+
 V_{M_f M_i}^{(1)}(\vec{q}\,',\vec{q})\,,
\end{equation}
which contains the Coulomb potential in momentum space, cut-off at radius R.
The so-called {\it first-order potential} of the strong pion-nuclear 
interaction ${\rm V}_{M_f M_i}^{(1)}$ is related to the free $\pi N$ 
scattering t-matrix
\begin{equation}
 V_{M_f M_i}^{(1)}(\vec{q}\,',\vec{q})=
-\frac{\sqrt{{\cal M}(q'){\cal M}(q)}}{2\pi}<\pi (\vec{q}\,'),f \mid
\sum^A_{j=1}\,t_{j}(\omega)\,\mid i,\pi (\vec{q})>\,,
\end{equation}

The free pion-nucleon scattering t-matrix is defined in the following way
\begin{mathletters}
\begin{equation}
t_{\pi N}\equiv<\vec{q}\,',\vec{p}\,'\mid t(\omega)\mid \vec{p},\vec{q}>=
-\frac{2\pi}{\sqrt{\mu (q',p')\mu (q,p)}}\,f_{\pi N}(\omega,\theta_{\pi}^*)\,
\delta(\vec{p}\,'+\vec{q}\,'-\vec{p}-\vec{q})\,,
\end{equation}
\begin{equation}
 f_{\pi N}(\omega,\theta_{\pi}^*)=
 {A_{0}+A_{T}\vec{t}\cdot\vec{\tau}+i\vec{\sigma}\cdot
[\hat{\vec{q}}_{i}\times\hat{\vec{q}}_{f}]
(A_{S}+A_{ST}\vec{t}\cdot\vec{\tau})}\,,
\end{equation}
\end{mathletters}
where $\mu (q,p)=E_{\pi}(q)E_{N}(p)/\omega$ is the pion-nucleon reduced mass,
$\vec{p}$ and $\vec{p}\,'$ are the nucleon momenta
in the initial and final states (in pion-nuclear c.m. system). 
In the pion-nucleon scattering
amplitude  $f_{\pi N}$, which is in general relativistically invariant, 
$\hat{\vec{q}}_{i}$ and $\hat{\vec{q}}_{f}$ are the unit vectors for the 
initial and final pion momenta in the $\pi N$ c.m. system.  The vectors
$\vec{\tau}$ and $\vec{t}$ are the usual isospin operators for the target
nucleon and pion.

The scalar functions $A_{i}(\omega,\cos{\theta_{\pi}^*})\,(i=0,S,T$, and
$ST$), that depend on the total pion-nucleon energy $\omega$ and the pion 
angle $\theta_{\pi}^*$ in the $\pi N$ c.m. system, are the usual combinations 
of partial $\pi N$ scattering amplitudes $f_{l_\pi}^{(\pm)}(\omega)$ and 
Legendre polynomials $P_{l_\pi}(\cos{\theta_{\pi}^*})$,  where $l_\pi$ is
the pion-nucleon angular momentum. The definition of these amplitudes
has been given in our previous work~\cite{KTB93}. 
Note that their on-shell value was obtained from  phase-shift analysis
of the elastic pion-nucleon scattering data.

The off-shell extrapolation of the $\pi N$ partial amplitudes was
constructed using a separable form
\begin{equation}
f_{l_\pi}^{(\pm)}({\rm {off-shell}})=f_{l_\pi}^{(\pm)}(\omega)
\frac{v_{l_\pi}^{(\pm)}(q'') v_{l_\pi}^{(\pm)}(q')}{[v_{l_\pi}^{(\pm)}(q)]^2}
\end{equation}
with $\pi N$ form factors
\begin{equation}
v_{l_\pi}^{(\pm)}(q) = \frac{q^{l_\pi}}{[1+(r_{0}q)^2]^2}\,\,.
\end{equation}
Here we employ the value $r_0 = 0.47 fm$, consistent with the analysis
of the separable $\pi N$ potential of Ref.\cite{Londer}
 
Finally, we mention a useful approximation
related to the treatment of nucleon Fermi motion. In accordance with the
results of Refs.\cite{Landau,TRD} the internal nucleon momentum can be replaced 
by an effective value
\begin{equation}
\vec{p}\rightarrow \vec{p}_{eff.}=-\frac{\vec{q}}{2} - 
\frac{1}{4}(\vec{q}-\vec{q}\,')\,.
\end{equation}
This choice is based on the fact that in the case of Gaussian wave
functions, which can reproduce the dominant S-wave part of the
deuteron ground state at low momenta, such a replacement treats the
linear $\vec{p}/2M$ terms in the pion-nucleon scattering amplitude
exactly.

\subsection{Partial wave decomposition and polarization observables.}
We express the scattering amplitude in terms
of partial amplitudes using  the representations of total angular momentum 
$j$ with projection $m$
\begin{equation}
F_{M_f M_i}(\vec{q}\,',\vec{q})=4\pi\sum (2j+1)i^{L_{\pi}-
L'_{\pi}}Y_{M'_{\pi}}^{L'_{\pi}}(\hat{\vec{q'}}) F^{j}_{L'_{\pi}L_{\pi}}
(q',q)Y_{M_{\pi}}^{*L_{\pi}}(\hat{\vec{q}})
\left({L'_{\pi}\;\;1\;\;\;\;\;\;\;j}
\atop{M'_{\pi}\,M_f\,-m}\right)
\left({L_{\pi}\;\;1\;\;\;\;\;\;j}
\atop{M_{\pi}\,M_i\,-m}\right)\,,
\end{equation}
where the $Y_{M_{\pi}}^{L_{\pi}}(\hat{\vec{q}})$
are the spherical harmonics for the
pion waves and $L_{\pi}\,(L'_{\pi})$ is the angular momentum of the
incoming (outgoing) pions. Note that due to parity conservation
$(-1)^{L_{\pi}+L'_{\pi}}=1$. For the amplitude ${\rm V}_{M_f M_i}$ we 
perform an expansion identical to Eq. (15).

Substituting the above expansions of $V_{M_f M_i}$ and
$F_{M_f M_i}$ into Eq. (8) we obtain the following system of integral 
equations for the partial wave amplitudes
\begin{equation}
F^{j}_{L'_{\pi}L_{\pi}}(q',q)=V^{j}_{L'_{\pi}L_{\pi}}(q',q)-\frac{a}{\pi} 
\sum_{{\tilde L}} \int \frac{q''^2}{{\cal{M}}(q'')}\frac{V^{j}_{L'_{\pi}
{\tilde L}}(q',q'') F^{j}_{{\tilde L}L_{\pi}}(q'',q)}{E(q) - E(q'') + 
i\epsilon} dq''\,.
\end{equation}
This equation is solved using the matrix inversion method. In our
evaluation of the partial amplitudes we also take into account
the Coulomb interaction applying the matching procedure developed by 
Vincent and Phatak~\cite{Vincent}.
 
Expressions for the polarization observables can be obtained using 
the spherical tensor operator $\tau_{k\kappa}$ which is defined as
\begin{equation}
 <J\,M'\mid \tau_{k\kappa} \mid J\,M> = {\hat J}{\hat k} (-1)^{J+M'}
\left({\,J\;\;\;\;\;\;\;J\;\;\;\;k}
\atop{\,M\;\;-M'\;\;\kappa}\right)\,,
\end{equation}
were we use the notation $\hat{J}=\sqrt{2J+1}$. Analyzing powers can
then be expressed as expectation values
\begin{equation}
T_{k\kappa}=\frac{Tr(F\tau_{k\kappa}F^{\dag})}{Tr(F\,F^{\dag})}\,\,.
\end{equation}

There are four polarization observables which can be measured in $\pi d$ 
elastic scattering using a pion beam and a polarized deuteron target. These 
are the vector analyzing power $iT_{11}$, and the three tensor analyzing powers
$T_{20}\,,T_{21}$ and $T_{22}$. The last two can be measured only in
linear combinations, such as
\begin{mathletters}
\begin{equation}
\tau_{21}=T_{21}+\frac{1}{2}(T_{22}+T_{20}/\sqrt{6})\,,
\end{equation}
\begin{equation}
\tau_{22}=T_{22}+T_{20}/\sqrt{6}\,.
\end{equation}
\end{mathletters}

The differential cross section and four polarization observables can be 
expressed in terms of five amplitudes, called
${\cal A},\,{\cal B},\,{\cal C},\,{\cal D}$ and ${\cal E}$,
which correspond to our amplitudes $F_{M_f M_i}$
with different initial and final deuteron spin projections. Following
{\it Robson's convention}~\cite{Robson} they are defined as
\begin{equation}
F_{M_f M_i} = \left( \begin{array}{ccc}
F_{++} & F_{0+} & F_{-+} \\
F_{+0} & F_{00} & F_{-0} \\
F_{+-} & F_{0-} & F_{--} \\
\end{array} \right) =
\left( \begin{array}{ccc}
\;{\cal A}    &\,{\cal B} &\,{\cal C} \\
 {\cal D} &\,{\cal E} &\,{\rm-}{\cal D} \\
\;{\cal C}    &\!{\rm-}{\cal B} &\;{\cal A} \\
\end{array} \right)\,,
\end{equation}
where the sign $+,0,-$ corresponds to the deuteron's spin projection
$M_{i(f)}=+1,0,-1$. Then the five observables can be written as
\begin{mathletters}
\begin{equation}
\frac{d\sigma}{d\Omega}=\frac{2q'}{3q}(\mid {\cal A} \mid^2+
\mid {\cal B} \mid^2 + \mid {\cal C} \mid^2+
\mid {\cal D} \mid^2 +\frac{1}{2} \mid {\cal E} \mid^2)
\equiv \frac{q'}{q}a(\theta_{\pi})\,,
\end{equation}
\begin{equation}
iT_{11}=\sqrt{\frac{2}{3}}\,Im\,[\,{\cal D}^*({\cal A}-
{\cal C})-{\cal B}^*{\cal E})\,]/a(\theta_{\pi})\,,
\end{equation}
\begin{equation}
T_{20}=\frac{\sqrt{2}}{3}(\mid {\cal A} \mid^2+
\,\mid {\cal B} \mid^2 + \mid {\cal C} \mid^2 -
2 \mid {\cal D} \mid^2 -\mid {\cal E} \mid^2)/a(\theta_{\pi})\,,
\end{equation}
\begin{equation}
T_{21}=-\sqrt{\frac{2}{3}}\,Re\,[\,{\cal D}^*\,({\cal A}-
{\cal C})+{\cal B}^*{\cal E})\,]/a(\theta_{\pi})\,, 
\end{equation}
\begin{equation}
T_{22}=\frac{1}{\sqrt{3}}\,[\,2\,Re\,(\,{\cal A}^*\,{\cal C})-
\mid {\cal B} \mid^2]/a(\theta_{\pi})\,. 
\end{equation}
\end{mathletters}
Note that due to the relation~\cite{Robson}
\begin{equation}
{\cal E}= ({\cal A}-{\cal C})-\sqrt{2}({\cal B}+{\cal D}) \cot{\theta_{\pi}}
\end{equation}
only four amplitudes are independent.

   In our  numerical calculations we will use a system of coordinates which
corresponds to the {\em Madison Convention}: A right-handed coordinate 
system in which the positive $z$-axis is along the beam direction (along 
the initial pion momentum $\vec q$) and the $y$-axis is along the vector 
$[{\vec q}\times{\vec q}\,']$.

\subsection{Pion-nuclear potential}

Here we discuss the first-order potential $V_{M_f M_i}^{(1)}$,  
defined in Eq. (10). It can be rewritten as
\begin{equation}
V_{M_f M_i}^{(1)}(\vec{q}\,',\vec{q}) = -2 \frac{\sqrt{{\cal M}(q')
{\cal M}(q)}}{2 \pi} \int d\vec{r}\,\Psi_{M_f}^{*}(\vec r)
\exp{(\frac{i{\vec Q}\cdot{\vec r}}{2})}\,t_{\pi N}(\omega;\vec{q}\,',\vec{q})
\,\Psi_{M_i}(\vec r)\,,
\end{equation}
where ${\vec Q}={\vec q}-\vec{q}\,'$ is the momentum transfer.

The deuteron wave function in this expression is given by
\begin{equation}
\Psi_{m}(\vec r)=-\sqrt{3}\sum_{l,m_l}\,(-1)^{l}\frac{U_l(r)}{r}
\left({l\;\;\;\;1\;\;\;\;\;1}\atop{m_l\;m_s\,-m}\right)
\;Y_{m_l}^{l}(\hat{\vec r})\,\chi_{m_s}^{1}\,,
\end{equation}
where $l=0$ and 2 denote the $S$- and $D$-components, respectively,
$\chi_{m_s}^{s}$ denotes the spin wave function for the two-nucleon 
system with total spin $s=1$ and projection $m_s$. After a
multipole decomposition  for the exponential,
 $\exp{(\frac{i{\vec Q}\cdot{\vec r}}{2})}$,
in Eq. (23) the pion-nuclear potential can be expressed via reduced
nuclear matrix elements $M_{SLJ}(Q)$ defined as
\begin{equation}
<f|\sum_{j=1}^2\,\left[\,Y^{L}(\hat{\vec r}_j)\otimes
\sigma_j^S\,\right]^J_M\,j_L(\frac{Qr_j}{2})\,|i>=(-1)^{1-M_f}
\left({\;\;1\;\;\;J\;\;\;\;1}\atop{{\rm -}M_f\;M\;M_i}\right)\,M_{SLJ}(Q)\,,
\end{equation}
where $j_L(z)$ is the spherical Bessel function, $\sigma^S=1$ and
$\vec\sigma$ for $S=0$ and 1, respectively.

Using standard techniques one finds
\begin{equation}
M_{SLJ}(Q)=3\sqrt{\frac{3(S+1)}{4\pi}}\sum_{l'l}\hat{l}\hat{l'}
\hat{L}\hat{J}\left({\,l\;\;\;l'\;\;\;L\,}\atop{0\;\;\;0\;\;\;0}\right)\,
\left\{ \begin{array}{ccc}
 1  &  l' &  1 \\
 1  &  l  &  1 \\
 J  &  L  &  S \\
\end{array} \right\}\, R_{l'l}^{(L)}(Q)\,.
\end{equation}
The radial integrals $R_{l'l}^{(L)}(Q)$ are defined as
\begin{equation}
R_{l'l}^{(L)}(Q)=\int_{}^{}d r\,U_{l'}(r)\,j_L(\frac{Qr}{2})\,U_l(r)
\end{equation}
and satisfy the normalization condition
\begin{equation}
R_{00}^{(0)}(Q)+R_{22}^{(0)}(Q)\stackrel{Q\rightarrow 0}{=}\,1\,.
\end{equation}
Explicit expressions for the matrix
elements $M_{SLJ}(Q)$ are given in the Appendix.

The radial wave functions $U_l(r)$ are parameterized\cite{Lacombe}
as a discrete superposition of Yukava-type terms
\begin{mathletters}
\begin{equation}
U_0(r)=\sum_{n=1}^N\;C_n\,\exp{(-m_n\,r)}\,,\qquad\qquad\qquad\qquad\qquad
(^3S_1-{\rm component})
\end{equation}
\begin{equation}
U_2(r)=\sum_{n=1}^N\;D_n\,\exp{(-m_n\,r)}(1+
\frac{3}{m_nr}+\frac{3}{m_n^2r^2})\,,\qquad(^3D_1-{\rm component})
\end{equation}
\end{mathletters} 
where the 13 coefficients $C_n$, $D_n$ and masses $m_n$ have been calculated
following the prescription given in Ref.\cite{Lacombe}.

In order to facilitate a qualitative understanding of the behavior of
the polarization observables we present expressions for the five 
amplitudes ${\cal A},...,{\cal E}$ in the {\it plane wave approximation 
(PWIA)}
(the second term in Eq.(8) is dropped). Taking into account  only 
contributions from the $S$-wave 
component and their interference with the $D$-wave component of the 
deuteron wave function (neglecting terms with $l'=l=2$), we obtain
\begin{mathletters}
\begin{equation}
{\cal A}\cong 2\,A_0\,W_0\,\left[R_{00}^{(0)}-\frac{1}{2\sqrt{2}}\left(1 - 
3\cos{\theta_{\pi}}\right)R_{02}^{(2)}\right]\,,
\end{equation}
\begin{equation}
{\cal B}\cong -\sin{\theta_{\pi}}\,\left[\sqrt{2}\,A_S\,W_S\,
\left(R_{00}^{(0)} +\frac{1}{\sqrt{2}}R_{02}^{(2)}\right) - 
3\,A_0\,W_0\,R_{02}^{(2)}\right]\,,
\end{equation}
\begin{equation}
{\cal C}\cong\sqrt{\frac{3}{2}}\left(1+\cos{\theta_{\pi}}\right)\,A_0\,W_0\,
R_{02}^{(2)}\,,
\end{equation}
\begin{equation}
{\cal D}\cong \sin{\theta_{\pi}}\,\left[\sqrt{2}\,A_S\,W_S\,
\left(R_{00}^{(0)} +\frac{1}{\sqrt{2}}R_{02}^{(2)}\right) + 
3\,A_0\,W_0\,R_{02}^{(2)}\right]\,,
\end{equation}
\begin{equation}
{\cal E}\cong 2\,A_0\,W_0\,\left[R_{00}^{(0)}+\frac{1}{\sqrt{2}}\left(1 - 
3\cos{\theta_{\pi}}\right)R_{02}^{(2)}\right]\,.
\end{equation}
\end{mathletters}
Here, $A_0$ and $A_S$ are the isoscalar scalar and spin-flip $\pi N$-scattering
amplitudes from Eq. (11), and $R_{l'l}^{(L)}$ are the radial integrals 
defined in Eq. (27), the coefficients $W_0$ and $W_S$ are
kinematical factors  arising from the Lorentz transformation of the
elementary amplitude from the $\pi N$ c.m. to the $\pi d$ c.m. system. 
They can easily be obtained from Eqs. (10-11) and the angle transformation of
$[\vec{q}_i\times\vec{q}_f]$ in the elementary amplitude~\cite{Landau}. 
The complete expressions of the five amplitudes
(including the full contribution from the deuteron $D$-state) are given
in the Appendix.

\section{ RESULTS AND DISCUSSION}
 
\subsection{Pion scattering on unpolarized  targets}
 
We begin our discussion with some of the main features of
the pion-nuclear interaction. The results of our calculations for pion
kinetic energies in the lab system, $T_{\pi}$= 65, 181 and 254 MeV,  are
shown in Fig. 1. One of the most important properties of the $\pi N$ 
interaction in this energy region is the dominance of the $p$-wave 
contribution coming from the $\Delta$-isobar excitation. This feature is 
reflected in the coherent scattering process which is proportional to $A$ 
(nuclear mass number) and described by the scalar-isoscalar
part ($A_{0}$) of the $\pi N$ amplitude. Since the $p$-wave part of this 
amplitude has a $\cos{\theta_{\pi}}$ dependence, the differential cross 
section experiences a minimum around $\theta_{\pi}=90^{\circ}$ (see dashed 
curves). The position of this minimum is slightly shifted from
the Lorentz transformation of the pion angle from the $\pi N$ to the $\pi d$ 
c.m. frame and from the $s$- and
$p$-wave interference. The spin-flip transition coming from the amplitude
$A_{S}$ (which is also mainly of $p$-wave nature), is proportional to
$\sin{\theta_{\pi}}$. The corresponding spin-flip contribution fills in the
minimum, as shown in Fig. 1.
 
In the $\Delta$-resonance region ($T_{\pi}$=180 MeV) our results are in 
fairly good agreement with the experimental data and are basically
the same as the results of three-body Faddeev calculations~\cite{Garc}.
This may be due to the dominance of the $\Delta$-resonance
contribution. It had been shown in Ref.\cite{Kopal} that in this case
the two-body approach is a good approximation for the more elaborate 
three-body Faddeev framework.

However, at lower energies ($T_{\pi}=65$ MeV) our model  fails to reproduce
the experimental data in the forward direction. This is related
to the well-known problem of describing the low-energy $S$-wave pion-nucleon
interaction in nuclei with zero isospin. In this case, the contribution from
the large isovector $\pi N$ scattering amplitude $A_T$ cancels and only
the small $A_0$ amplitude remains in the first-order potential. Therefore, 
in the zero-energy limit our approach gives a small value for
the $\pi d$-scattering length
\begin{equation}
a_{\pi d}^{(1)}=\lim_{q\rightarrow 0}\frac{\tan{\delta_0^{(1)}}}{q}
\approx -0.015 fm\,,
\end{equation}
where $\delta_0^{(1)}$ is the $S$-wave $\pi d$ scattering phase shift 
calculated with the first-order potential. The experimental value is about
five times larger: $a_{\pi d}^{exp.}=-(0.073\pm 0.02) fm$~\cite{apid}.
In such a situation one expects higher-order effects in the pion-nuclear
interaction to be important. The well-known result (see for example 
Refs.\cite{Koltun,Ericson}) for the second-order
correction is
\begin{equation}
a_{\pi d}^{(2)}=2\,(A_0^2-2A_T^2)\,\left<\frac{1}{r}\right>\,C^2\,,
\end{equation}
where $<1/r>=0.46 fm^{-1}$ and $C=(1+m_{\pi}/M)(1+m_{\pi}/2M)^{-1}$.
In this expression the contribution from the isovector amplitude $A_T$  which 
describes double isovector scattering (including charge exchange) becomes
important. Taking into account such a correction for the scattering length we
obtain 
\begin{equation}
a_{\pi d}\approx -(0.015+0.034)fm =-0.049 fm
\end{equation}
which is in better agreement with the experimental value as well as
with results from Faddeev calculations: $a_{\pi d}^{Fad.}=0.046 fm$
\cite{Petrov} (for the same set of $A_0$ and $A_T$ amplitudes).
As shown in Fig. 1, this correction factor also improves the agreement with the
experimental data for the differential cross section at $T_{\pi}$=65 MeV
(see dash-dotted curve).
 
The role of the $^3D_1$-configuration in the deuteron wave function is
also illustrated in Fig. 1. Its contribution can be seen by
turning off the nuclear matrix elements of Eq. (26) with $L=2$ (see dotted
curves in Fig. 1), thus retaining the normalization condition of Eq. (28).
As expected, the $D$-state contribution becomes significant at
large angles, $\theta_{\pi}>90^0$,  for pion energies
$T_{\pi}>$180 MeV which correspond to momentum transfers of
 $Q > 1.78 fm^{-1}$.

Fig. 2 presents our results for higher pion energies. Note that with
increasing pion energies the 
second-order correction discussed above
becomes less important due to the dominance of the pion-nucleon $p$-wave 
contribution. However, it is still visible at $T_{\pi}$=142 MeV and it
improves the description in the forward direction (the corresponding result 
is not shown in Fig. 2). Fig. 2 illustrates the role of pion rescattering
contributions coming from the second term of Eq. (8) by comparing the plane 
wave (without second term of Eq. (8))
calculations (dashed curves)
with the full results (solid curves). Pion rescattering becomes very important
in the $\Delta$-region ($140<T_{\pi}<250$ MeV), especially at large angles.
Here it can reduce the plane wave results by a factor of two.
However, with further increase of energy the $\pi N$-interaction
becomes weaker again, and, consequently, the rescattering contributions become
smaller.

At $T_{\pi}>250$ MeV the main problem which arises in our approach
(as well as Faddeev calculations) is the discrepancy with experimental data at
backward angles. In contrast to Faddeev results which usually
overestimate the data in this region by a factor of two~\cite{Garc},
our calculations tend to underestimate the measurements.  We point out that 
a similar disagreement has been found for pion scattering on 
$^3$He and $^4$He~\cite{KTB93,He4}.

In the framework of Faddeev equations there have been numerous attempts to
improve the situation at backward angles, mainly by changing the
two-body inputs. One of the more controversial attempts addressed the 
$P_{11}$ part of the $\pi N$ amplitude with its division into a pole and 
nonpole term, summarized in Refs.\cite{Garc,Greg}. We only note here that 
the pole term determines the coupling between the $NN$ and $\pi NN$ channels.
Due to this coupling, pion scattering can also be presented as pion absorption 
on one nucleon and subsequent emission from another. Such a two-body mechanism 
is not included in the standard multiple scattering theory. At large momentum 
transfers this process could be important because of the possibility 
of momentum sharing. However, as argued by Jennings~\cite{Jen}, such a 
two-body mechanism is almost completely canceled by
additional contributions which come from the  time
ordering of the pion absorption and emission processes.

On the other hand, the issue concerning the
so-called $P_{11}$-problem cannot be regarded as settled since
the strong cancellation takes place
only if there is no interaction between the two nucleons in the intermediate
state. Our simple estimates show that if we would include
the repulsive part of the $NN$-interaction which involves
heavy meson exchange (as was done by Mizutani and Koltun in
their study of the $\pi d$ scattering length~\cite{MK77}), the cancellation
disappears.
After a modification of the free two-nucleon propagator by adding   
the mean value of the residual $NN$-interaction $<V_{NN}(\rho,\omega,...)>$ 
the corresponding two-body contribution becomes proportional to $<V_{NN}>$. 
Fig. 2 illustrates the effects which could come from this two-body mechanism 
with $<V_{NN}>=40$ MeV. Clearly, a more microscopic calculation is called for, 
here we want to merely point out that this mechanism can improve
the situation at backward angles.
 
\subsection{Polarization observables.}
 
Here we begin our discussion with the results for the tensor analyzing powers.
In order to understand the general structure of these observables and their 
behavior in different kinematical regions we first consider the size of
the observables at backward angles $\theta_{\pi}=180^0$.
It is instructive to also neglect the
contribution of the deuteron $^3D_1$-component. As an
example, Fig. 3 presents $T_{20}$, $\tau_{21}$ and $\tau_{22}$
at $T_{\pi}=294$ MeV. We remind the reader that, according to Eq. (19), the last
two observables are linear combinations of $T_{20}$, $T_{21}$ and
$T_{22}$.

It follows from Eq. (30) that at $\theta_{\pi}=180^0$ the amplitudes
$\cal B$, $\cal C$ and $\cal D$ vanish. Therefore, at backward angles
the observables $T_{21}$ and $T_{22}$ vanish as well, and  $T_{20}$ reduces
to the following expression
\begin{equation}
T_{20}(180^0)=\frac{\sqrt{2}(\mid {\cal A}\mid^2-\mid {\cal E}\mid^2)}
{2\mid {\cal A}\mid^2+\mid {\cal E}\mid^2}\,.
\end{equation}
From Eq. (30) we can see that in the region where the radial integral
$R_{00}^{(0)}=\sqrt{2}R_{02}^{(2)}$, the contribution from the $\cal A$
amplitude becomes very small. In this region we then have
\begin{equation}
T_{20}(180^0)\approx-\sqrt{2}\,,\qquad \tau_{22}(180^0)\approx
2\tau_{21}(180^0)\approx-\frac{1}{\sqrt{3}}\,.
\end{equation}
This result is confirmed experimentally for $T_{\pi}>200$ MeV.
Note, that at lower energies  in the region where the deuteron $D$-wave
contribution is small ($R_{00}^{(0)}>>R_{02}^{(2)}$) tensor 
observables at backward angles become proportional to the strength of 
the deuteron $D$-state. For example,
$T_{20}(180^0)\approx -4 R_{ 02}^{(2)}/R_{00}^{(0)}$.

Another simplified scenario is given by neglecting
the pion rescattering contributions and
the deuteron $D$-wave. In this case, the $\cal C$-amplitude vanishes,
and ${\cal A}^{(S)}={\cal E}^{(S)}$ and  ${\cal B}^{(S)}=-{\cal D}^{(S)}$.
Due to these identities, $T_{21}^{(S)}=0$ and the other
observables can be expressed solely via the
elementary $\pi N$ scattering amplitudes
\begin{equation}
T_{20}^{(S)}=-\frac{2\sqrt{2}\mid A_S \mid^2 \sin^2{\theta_{\pi}}}
{12 \mid A_0 \mid^2 + 8\mid A_S \mid^2\,\sin^2{\theta_{\pi}}} ,
\qquad T_{22}^{(S)}=\sqrt{\frac{3}{2}}T_{20}^{(S)}\,.
\end{equation}
As we have shown in the analysis of the differential cross sections, due to
the dominance of the  $p$-wave contribution in $\pi N$ scattering,
$A_0(90^0)\approx 0$. Therefore, in the $\theta_{\pi}=90^0$ region where 
the contribution from the deuteron $S$-state dominates we have
\begin{equation}
T_{20}^{(S)}(90^0)\approx -\sqrt{2}/4\,,\qquad
\tau_{22}^{(S)}(90^0)\approx 2\,\tau_{21}^{(S)}(90^0)\approx 
-\frac{1}{\sqrt{3}}\,.
\end{equation}
Fig. 3 shows that these predictions  can already reproduce the experimental 
measurements fairly well.

The next three figures, Fig. 4, Fig. 5 and Fig. 6, present tensor
observables at $T_{\pi}$=134, 180, 219 and 256 MeV. In general, their
behavior follows the simple picture described above.
At backward angles $T_{20}$ and $T_{22}$ are
almost completely given by the $^3D_1$-component of the deuteron wave function.
In forward direction (up to $\theta_{\pi}=100^0$) the main contribution
comes from the $^3S_1$-component. Here $T_{20}$ and $T_{22}$  contain very
little nuclear structure information.  $T_{21}$ depends entirely on the
deuteron $D$-state.  In the case of the $\tau_{21}$-observable,
which is mainly associated with $T_{21}$, the deuteron $S$-state tends 
to fill in the minimum around $\theta_{\pi}=90^0$. Note that this minimum 
appears because of the p-wave dominance of the $\pi N$ scattering amplitude.
Finally, we see that the influence of pion rescattering effects on
all tensor analyzing powers  is small.

Thus, we can obtain a good description for the tensor observables.
This may be related to the fact that they are determined mainly by the 
absolute values of the ${\cal A}-{\cal E}$ amplitudes. In contrast, 
the vector analyzing power $iT_{11}$ depends on the interference between 
these amplitudes. Therefore, we expect this quantity to be more sensitive 
to details of the model ingredients. In Fig. 7 we compare our results with 
some old~\cite{Smith} and more recent~\cite{Wessler} measurements.
We obtain satisfactory agreement with these data only at $T_{\pi}$=100 MeV. 
At all other energies our results fail to reproduce
the experimental data, especially for $\theta_{\pi}>90^0$. The failure
becomes more dramatic with increasing pion energy. In our
full calculation $iT_{11}$ goes through zero around $T_{\pi}$=180 MeV 
and becomes negative at $\theta_{\pi}>90^0$. We have found that pion 
rescattering is mainly responsible for this effect. Experimental
measurements show quite the opposite behavior: Above $T_{\pi}$=250 MeV,
$iT_{11}$ changes the sign in forward direction producing a negative
dip around $\theta_{\pi}=70^0$ and it assumes large positive values for
$\theta_{\pi}>90^0$. This behavior could only be described
in simple calculations without pion rescattering (dashed curves). Note that
the influence of the deuteron $D$-state on $iT_{11}$ is small. Therefore,
in the plane wave  approximation
the vector analyzing power can be expressed  via the elementary $\pi N$ 
scattering amplitudes
\begin{equation}
iT_{11}^{(S)}=-\frac{\sqrt{6} Im ( A_S^*\,A_0 )\, \sin{\theta_{\pi}}}
{12 \mid A_0 \mid^2 + 8\mid A_S \mid^2\,\sin^2{\theta_{\pi}}}\,.
\end{equation}
 
In this paper we have not attempted to present a detailed
comparison with the Faddeev calculations that have been performed.
A review of these results can be found in Ref.\cite{Garc}. We only mention 
that all conventional calculations encounter the difficulties in the 
description of the measured vector analyzing power in the $\Delta$-resonance 
region.

On a phenomenological level we have studied the effect of
modifying the spin-flip amplitude $A_S$ (similar to $A_0$ at low energies)
\begin{equation}
A_S\rightarrow A_S - B_S\,
\frac{<d\mid\,e^{i(\vec{q}+\vec{q'})\cdot\vec r/2}
\frac{e^{-m_{\pi}r}}{r}\,\mid d>}{<d\mid\,e^{i(\vec{q}-\vec{q'})\cdot\vec 
r/2}\mid d>}\,,
\end{equation}
where $\mid d>$ is the deuteron ground state.
The  phenomenological term with a complex parameter $B_S$ would be
associated with second-order contributions. 
The sensitivity of  the differential cross section and tensor
analyzing powers to this correction is small since they are dominated by the
coherent contribution from the non-spin flip amplitude $A_0$.
However, it has a dramatic influence on the vector analyzing power. 
Using this sensitivity we extracted the $B_S$ parameter from the experimental 
data for $iT_{11}$. The result of our fit is shown in Fig. 7 by the
dash-dotted curves. It is remarkable that the energy behavior of the real 
and imaginary parts of $B_S$ follows  a typical resonance structure 
(see Fig. 8). This might be an indication for the presence of a residual
$\Delta N$ interaction. We point out that a similar situation occurs
in pion scattering on $^3$He for the target asymmetry
$A_y=\sqrt{2}\,iT_{11}$~\cite{Dehn}. It seems reasonable to assume
that the origin of this phenomenon is similar for both reactions.
It therefore becomes very important to find  a consistent description of
polarization observables in pion scattering on the deuteron and $^3$He.   
 
\section{CONCLUSION}

In this paper, we have studied the interaction of pions with the
deuteron in a multiple scattering approach carried out in momentum space. 
Our investigation covered the energy region of $T_{\pi}$= 60 - 300 MeV, 
thus covering both the low-energy and the $\Delta$ region.
Paris wave functions were employed to describe the deuteron. Phase-shift
parameterizations were used for the elementary $\pi N$ amplitudes, along with
a separable potential for the off-shell extrapolation. The full spin
and isospin dependence of the $\pi N$ amplitudes were taken into
account.

Our framework of multiple scattering theory is clearly less
sophisticated than the many Faddeev calculations that have studied pion
deuteron scattering over the years. The primary purpose of our
investigation was to demonstrate that the approach developed 
recently for the description of the pion interaction with $A\geq 3$ nuclei 
can be reliably applied in the case of the deuteron. The results
obtained are transparent and can in the following be employed for the
study of pion photo- and electroproduction off deuterium. 

We found that our approach gives a good description of the differential 
cross sections and the tensor analyzing powers. At low pion energies, 
second-order corrections must be included, while at higher energies and 
backward angles discrepancies appear that may be related to two-body processes
involving heavy meson exchange. The tensor analyzing powers strongly
depend on the deuteron $D$-state and can be reproduced satisfactorily.
Clearly, the outstanding problem is the explanation of the
well-measured vector analyzing power which has defied a number of
attempts using more advanced approaches than ours. While there may be a
sensitivity regarding a residual $\Delta N$-interaction care has to
be taken to include it consistently. 

In the following paper (Part II),
we will show that the pion-deuteron interaction developed here
is adequate to describe the final state interaction in coherent pion 
photoproduction from deuterium.

\acknowledgements

We would like to thank Prof. E. Boschitz for helpful
discussions and providing us with the new data for the polarization 
observables. One of us (S.S.K) is grateful to the theory group of 
Prof. D. Drechsel for the hospitality extended to him during his stay at 
the University of Mainz. This work was supported by the Deutsche 
Forschungsgemeinschaft (SFB 201), the U.S. DOE grant 
DE-FG02-95-ER40907 and the Heisenberg-Landau program.

\appendix
\section*{}

In this Appendix, we give the complete expressions for the {\it Robson
amplitudes} ${\cal A},...,{\cal E}$ where the full contributions from the
deuteron $D$-state have been taken into account.

First, we express the nuclear matrix elements $M_{SLJ}(Q)$ defined
by Eqs. (25-26) via the radial integrals of Eq. (27)
\begin{equation}
M_{000}(Q)=2\,\sqrt{\frac{3}{4\pi}}\left[R_{00}^{(0)}(Q)+
R_{22}^{(0)}(Q)\right]\,,
\end{equation}
\begin{equation}
M_{022}(Q)=4\,\sqrt{\frac{3}{4\pi}}\left[R_{02}^{(2)}(Q)-
\frac{\sqrt{2}}{4}R_{22}^{(2)}(Q)\right]\,,
\end{equation}
\begin{equation}
M_{101}(Q)=2\,\sqrt{\frac{6}{4\pi}}\left[R_{00}^{(0)}(Q)-
\frac{1}{2}R_{22}^{(0)}(Q)\right]\,,
\end{equation}
\begin{equation}
M_{121}(Q)=-2\,\sqrt{\frac{6}{4\pi}}\left[R_{02}^{(2)}(Q)+
\frac{1}{\sqrt{2}}R_{22}^{(2)}(Q)\right]\,.
\end{equation}

 Then the  amplitudes ${\cal A},...,{\cal E}$ in the PWIA approach can
 be written as
\begin{equation}
{\cal A}= \sqrt{\frac{4\pi}{3}}\,A_0\,W_0\,\left[M_{000}(Q)-
\frac{1}{4\sqrt{2}}(1 - 3\cos{\theta_{\pi}})M_{022}(Q)\right]\,,
\end{equation}
\begin{equation}
{\cal B}= -\frac{1}{2}\sqrt{\frac{4\pi}{3}}\sin{\theta_{\pi}}\,
\left[A_S\,W_S\,M_S(Q)-\frac{3}{2}\,A_0\,W_S\,M_{022}(Q)\right]\,,
\end{equation}
\begin{equation}
{\cal C}=-\frac{3}{4}\sqrt{\frac{4\pi}{6}}\left(1+\cos{\theta_{\pi}}\right)
\,A_0\,W_0\,M_{022}(Q)\,,
\end{equation}
\begin{equation}
{\cal D}= \frac{1}{2}\sqrt{\frac{4\pi}{3}}\sin{\theta_{\pi}}\,
\left[A_S\,W_S\,M_S(Q)+\frac{3}{2}\,A_0\,W_S\,M_{022}(Q)\right]\,,
\end{equation}
\begin{equation}
{\cal E}= \sqrt{\frac{4\pi}{3}}\,A_0\,W_0\,\left[M_{000}(Q)+
\frac{1}{2\sqrt{2}}(1 - 3\cos{\theta_{\pi}})M_{022}(Q)\right]\,.
\end{equation}

In Eqs. (A6) and (A8) we introduced the matrix element
\begin{equation}
M_S(Q)=M_{101}(Q)-\frac{1}{\sqrt{2}}M_{121}(Q)\,.
\end{equation}

\begin{figure}
\caption{
Differential cross sections for $\pi^+$ 
elastic scattering on the deuteron at pion kinetic energies $T_{\pi}$=65, 
181 and 254 MeV calculated with (solid curves) and without (dashed curve)
spin flip transition. The dotted curves at $T_{\pi}$=181 and 254 MeV are the 
results obtained without the deuteron $D$-state. The dash-dotted curve at 
$T_{\pi}$=65 MeV is the result with second order corrections (32). 
Experimental data are from Refs. \protect\cite{Bal83,Gab80}($\bullet$) and 
Ref. \protect\cite{Otter1}(o).}
\end{figure}
 
\begin{figure}
\caption{
Differential cross sections for $\pi^+$ elastic scattering on
the deuteron at pion kinetic energies $T_{\pi}$=142--324 MeV. 
Solid and dashed curves are full and PWIA 
calculations, respectively. The dotted curves are our best fit with
a phenomenological interaction of two intermediate nucleons in the Jennings
mechanism. Experimental data are from 
Ref. \protect\cite{Gab80}($\bullet$), Ref. \protect\cite{Otter1}(o) 
and Ref. \protect\cite{Cole}($\triangle$).}
\end{figure}
 
\begin{figure}
\caption{
Tensor analyzing powers $T_{20},\,
\tau_{21}=T_{21}+\frac{1}{2}(T_{22}+T_{20}/\protect\sqrt{6})$ and 
$\tau_{22}=T_{22}+T_{20}/\protect\sqrt{6}$ at $T_{\pi}$=294 MeV (solid curves). 
Individual contributions from $T_{21}$ (in $\tau_{21}$)  and $T_{22}$ 
(in $\tau_{22}$) are shown by dashed 
curves. The dotted curves are the results obtained for $T_{20},\,\tau_{21}$ 
and $\tau_{22}$ without the deuteron $D$-state. Experimental data are from 
Ref. \protect\cite{Otter2}($\bullet$), Ref. 
\protect\cite{Otter3}(black triangles) and Ref. \protect\cite{Wessler}(o).}
\end{figure}

\begin{figure}
\caption{
$T_{20}$ observable at $T_{\pi}$=134--256 MeV. 
Solid and dashed curves are full and PWIA  
calculations, respectively. The dotted curves are the results obtained without
the deuteron $D$-state. Experimental data are from 
Ref. \protect\cite{Smith}($\bullet$) and Ref. \protect\cite{Otter2}(o).}
\end{figure}
 
\begin{figure}
\caption{
The same as in Fig. 4 for the observable 
$\tau_{21}=T_{21}+\frac{1}{2}(T_{22}+T_{20}/\protect\sqrt{6})$.
Experimental data are from Ref. \protect\cite{Smith}($\bullet$)
and Ref. \protect\cite{Otter2}(o).}
\end{figure}
 
\begin{figure}
\caption{
The same as in Fig. 4 for the observable 
$\tau_{22}=T_{22}+T_{20}/\protect\sqrt{6}$.
Experimental data are from Ref. \protect\cite{Otter3}($\bullet$) and 
Ref. \protect\cite{Wessler}(o).}
\end{figure}
 
\begin{figure}
\caption{
Vector analyzing power $iT_{11}$ at {\bf a)} $T_{\pi}$=100--164 MeV 
{\bf b)} $T_{\pi}$=180--294 MeV.
Solid and dashed curves are full and PWIA 
calculations, respectively.  Dash-dotted curve are the results of our fit with 
phenomenological term (39). Experimental data are from 
Ref. \protect\cite{Smith} ($\bullet$) and Ref. \protect\cite{Wessler}(o).}
\end{figure}
 
\begin{figure}
\caption{
Energy dependence of the modification for the spin-flip interaction
(see  Eq. (39)) extracted from the vector analyzing power $iT_{11}$.}
\end{figure}
 
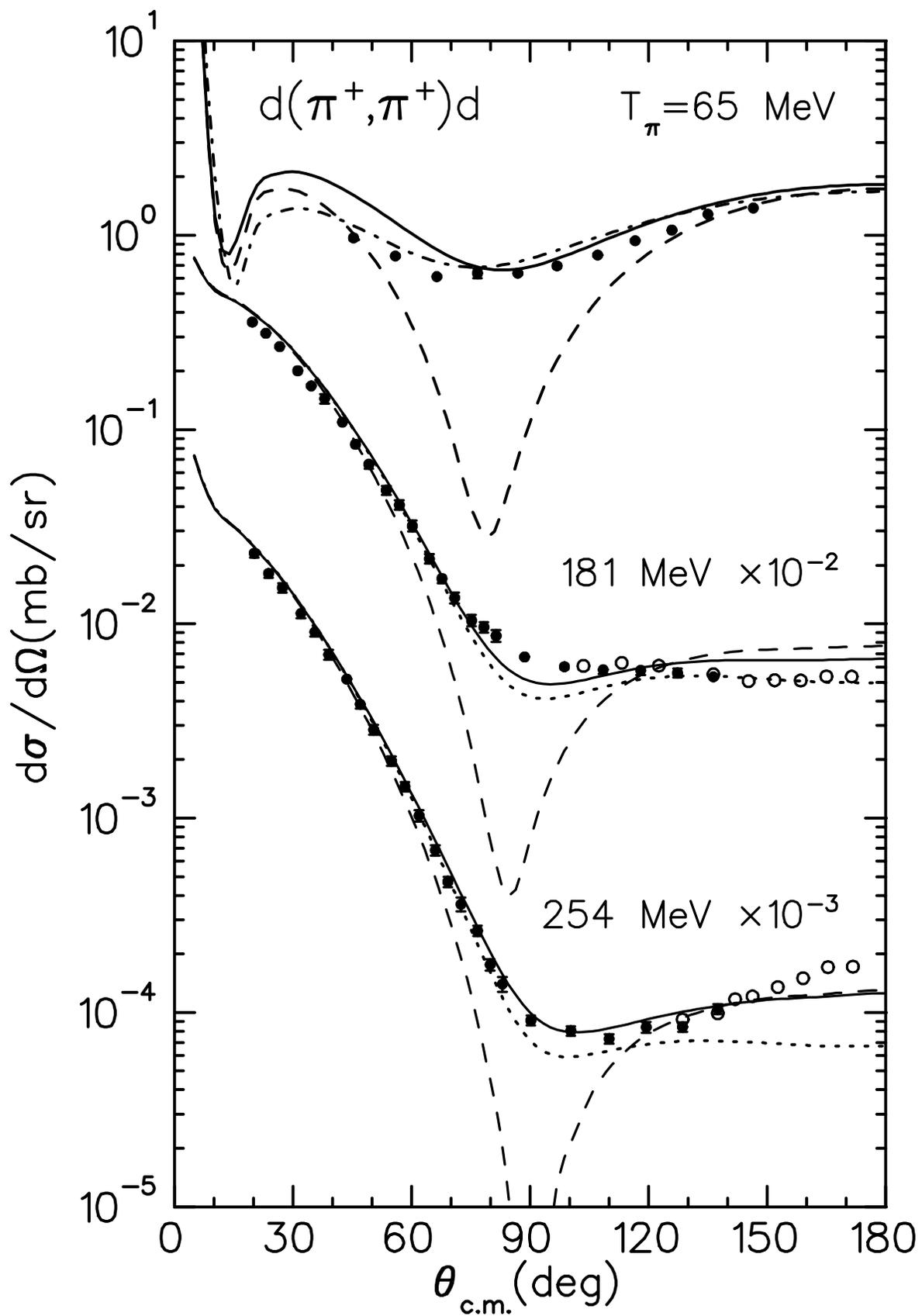
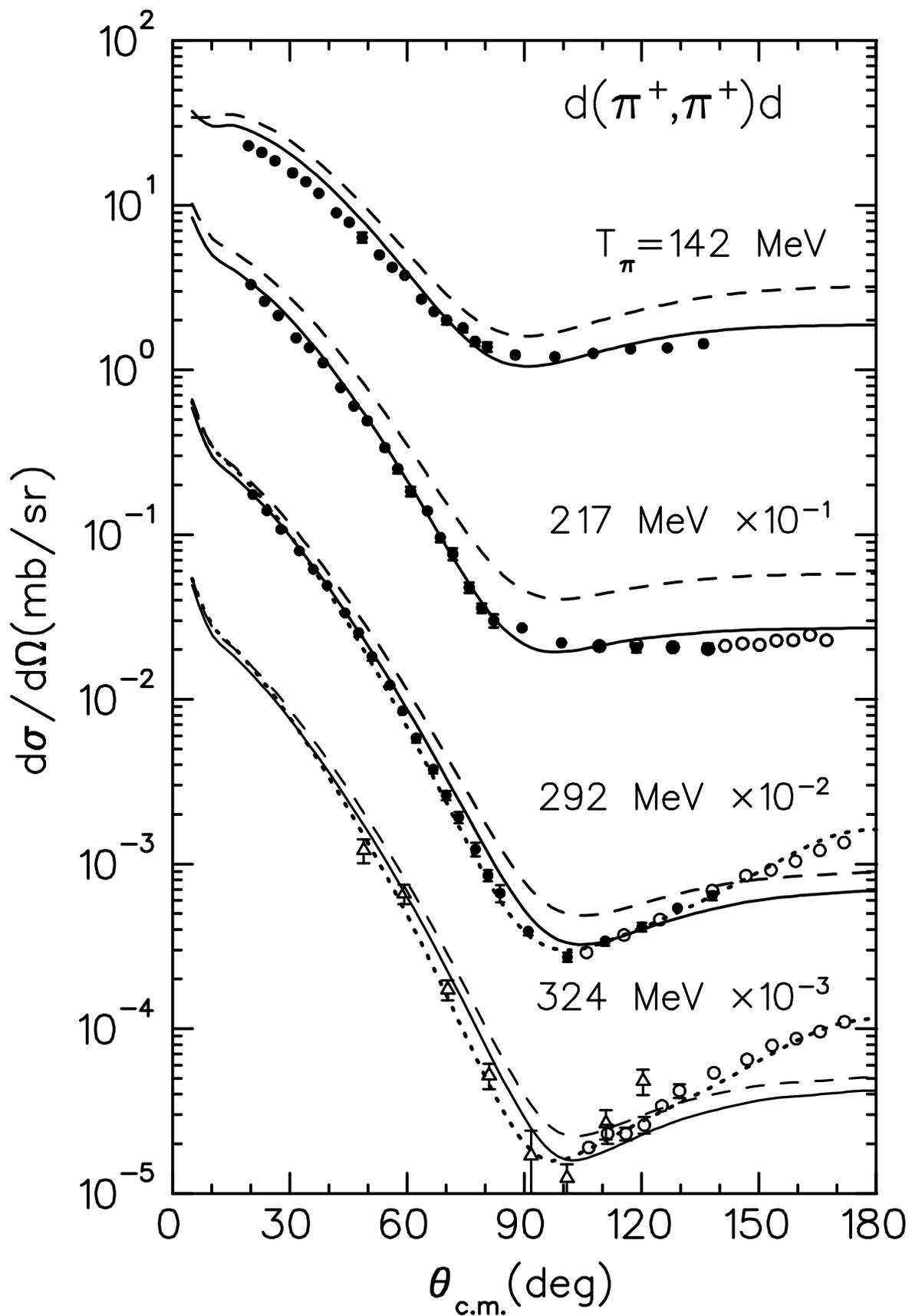
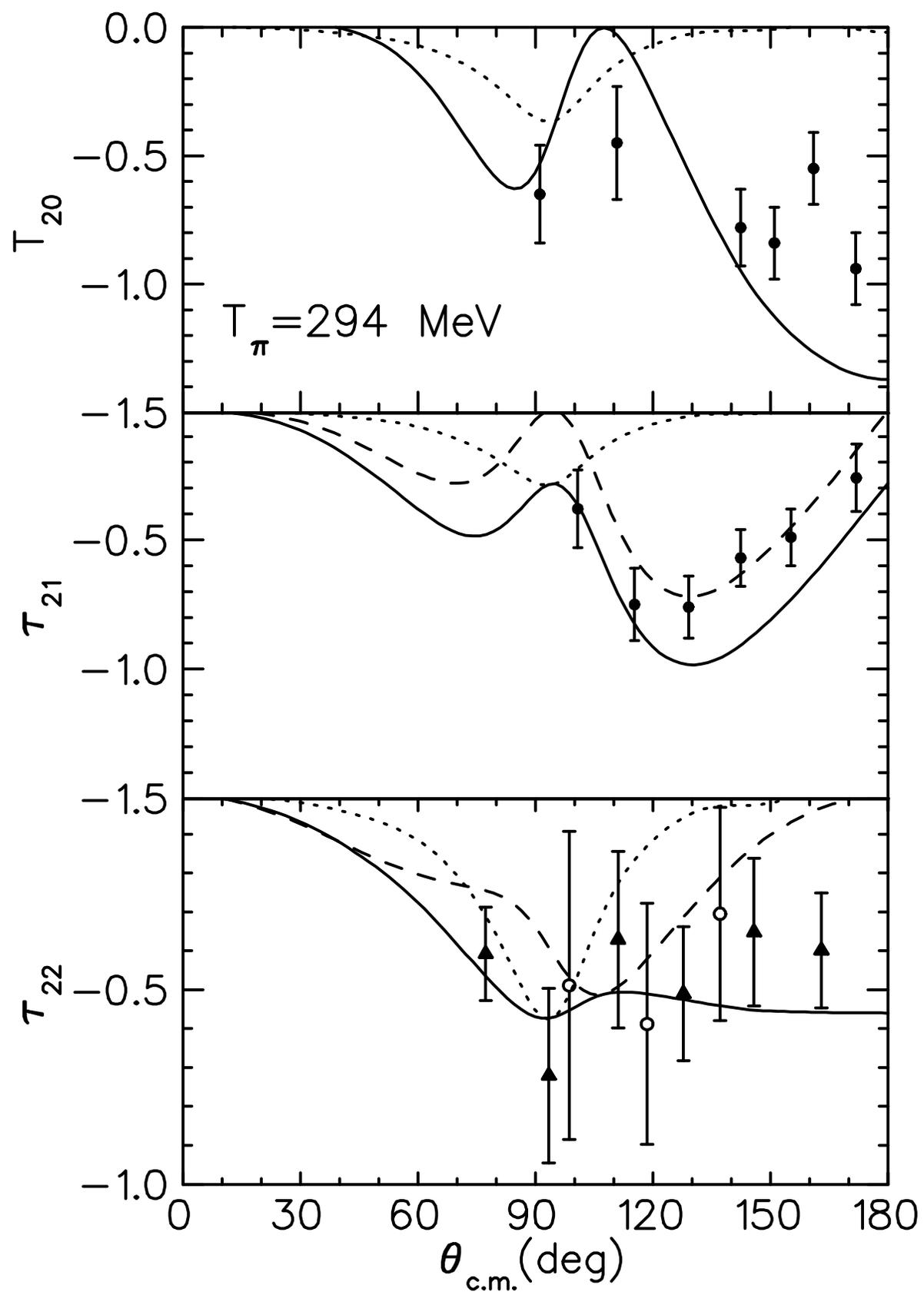
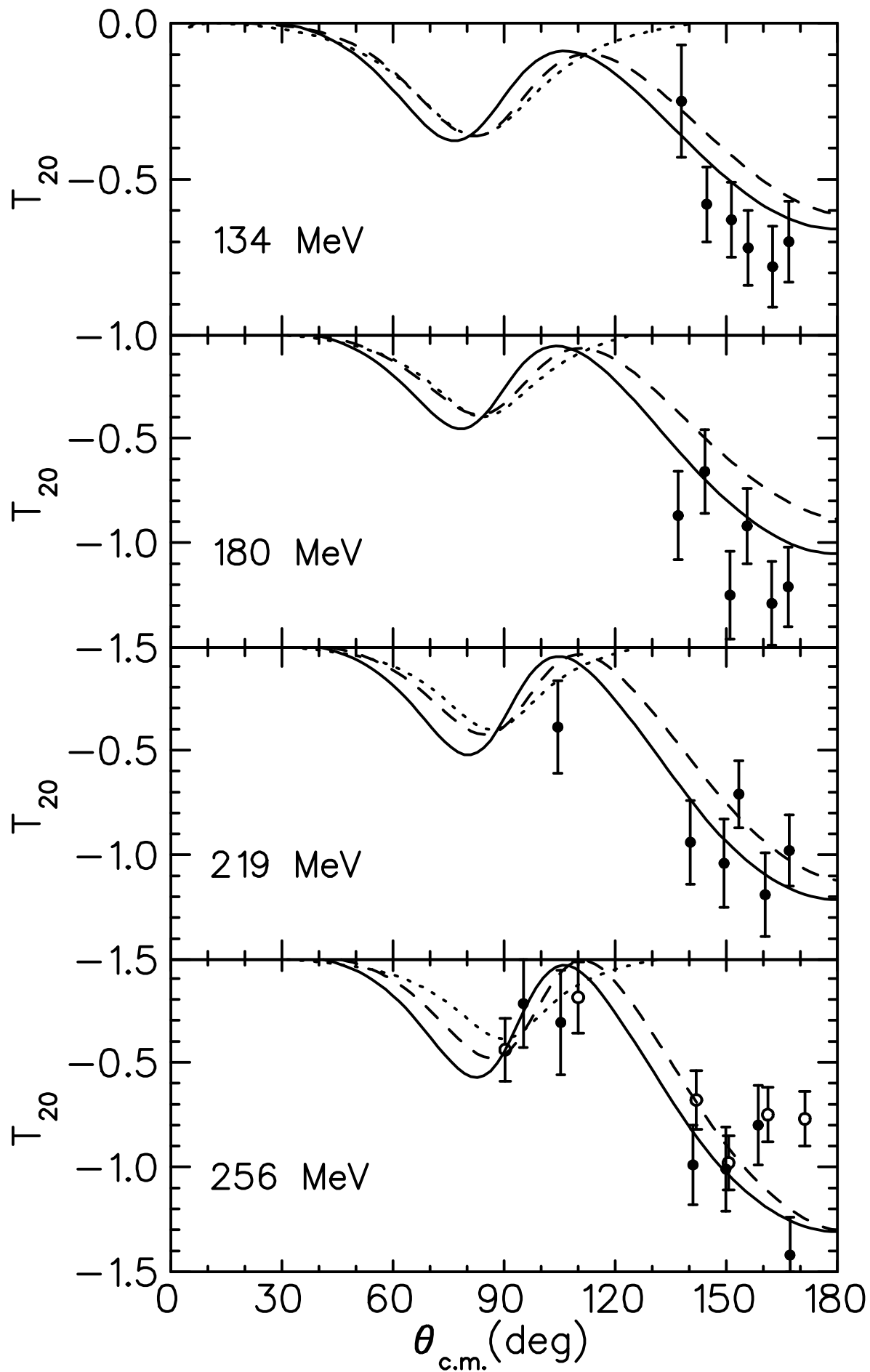
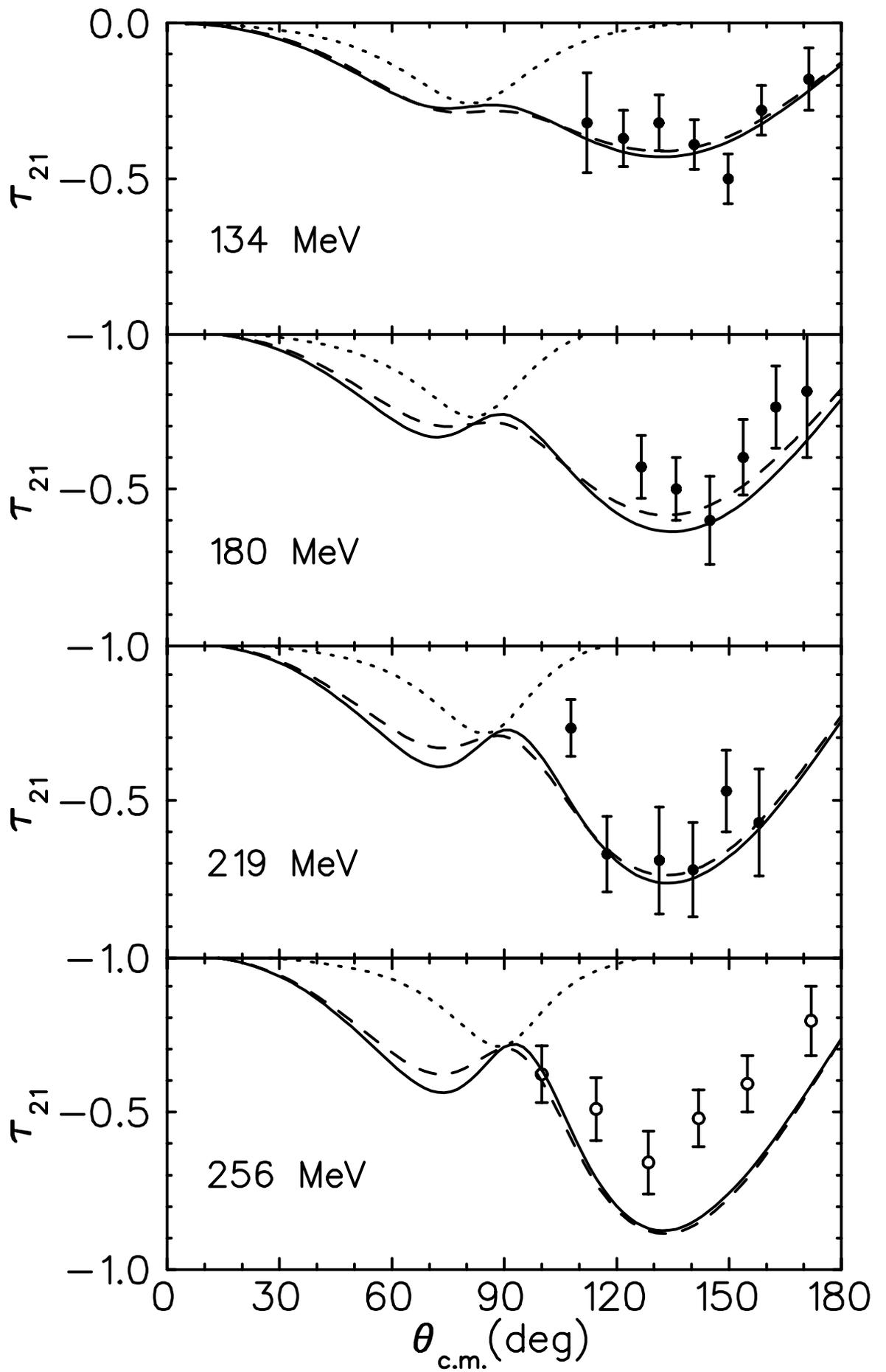
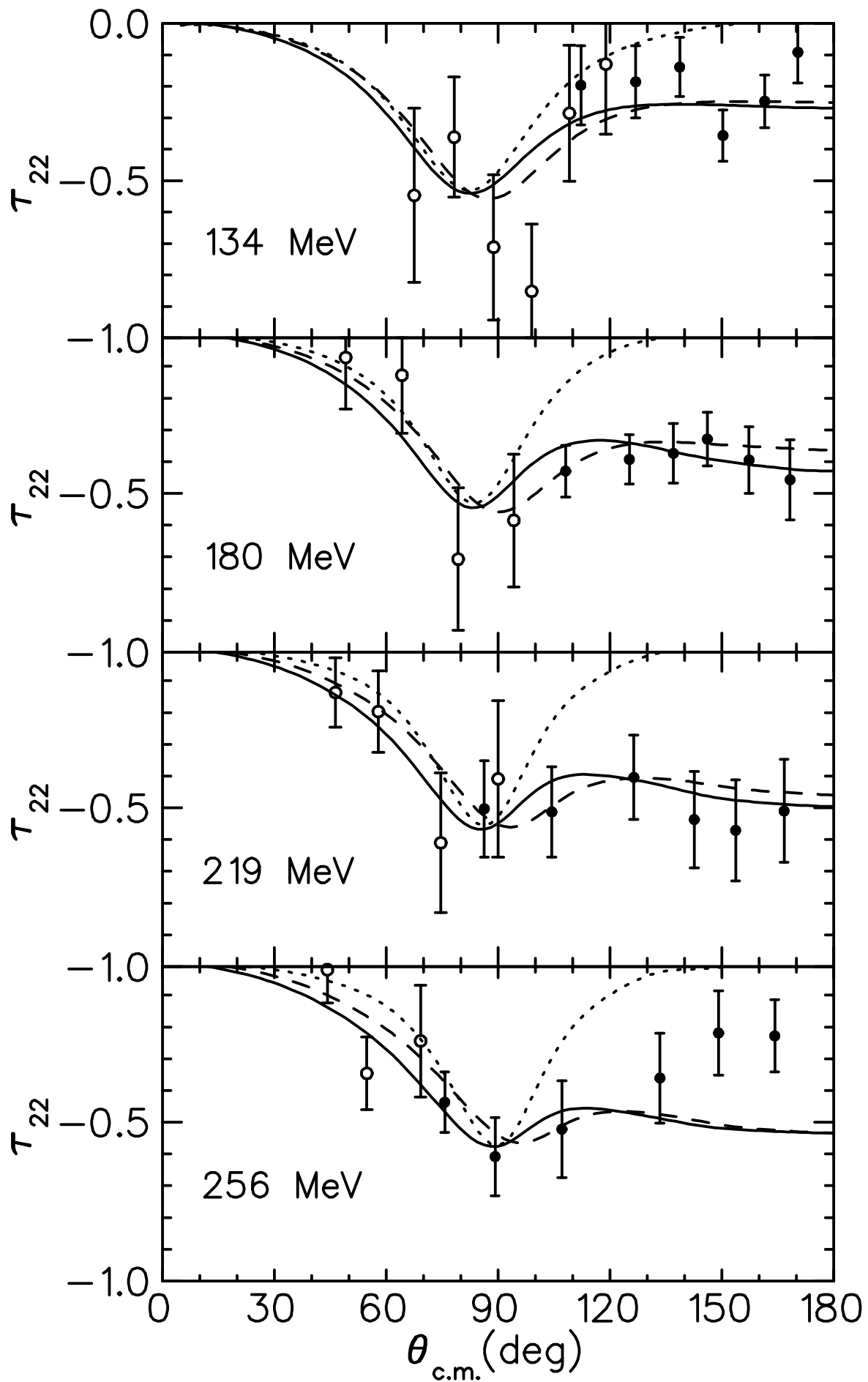
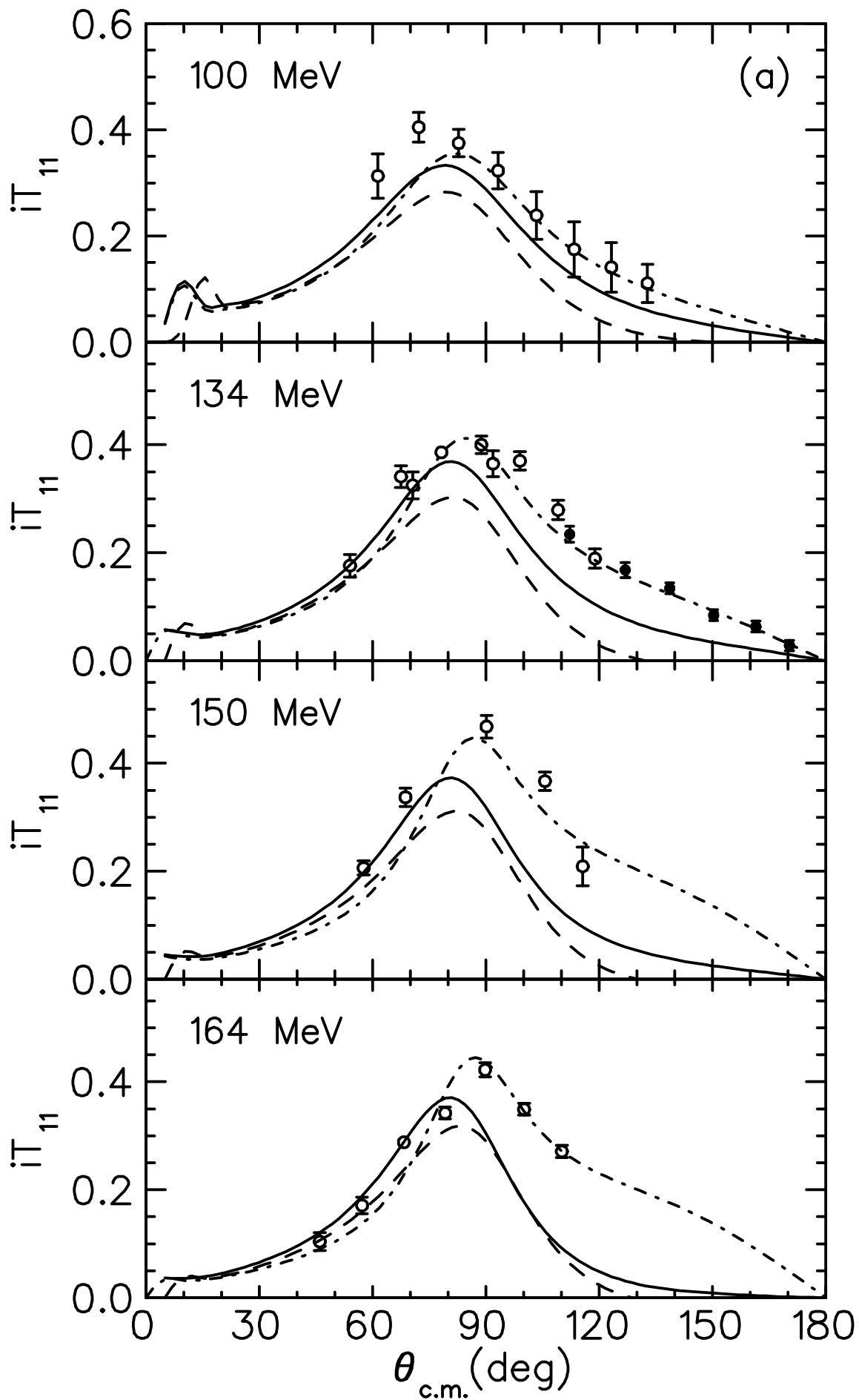
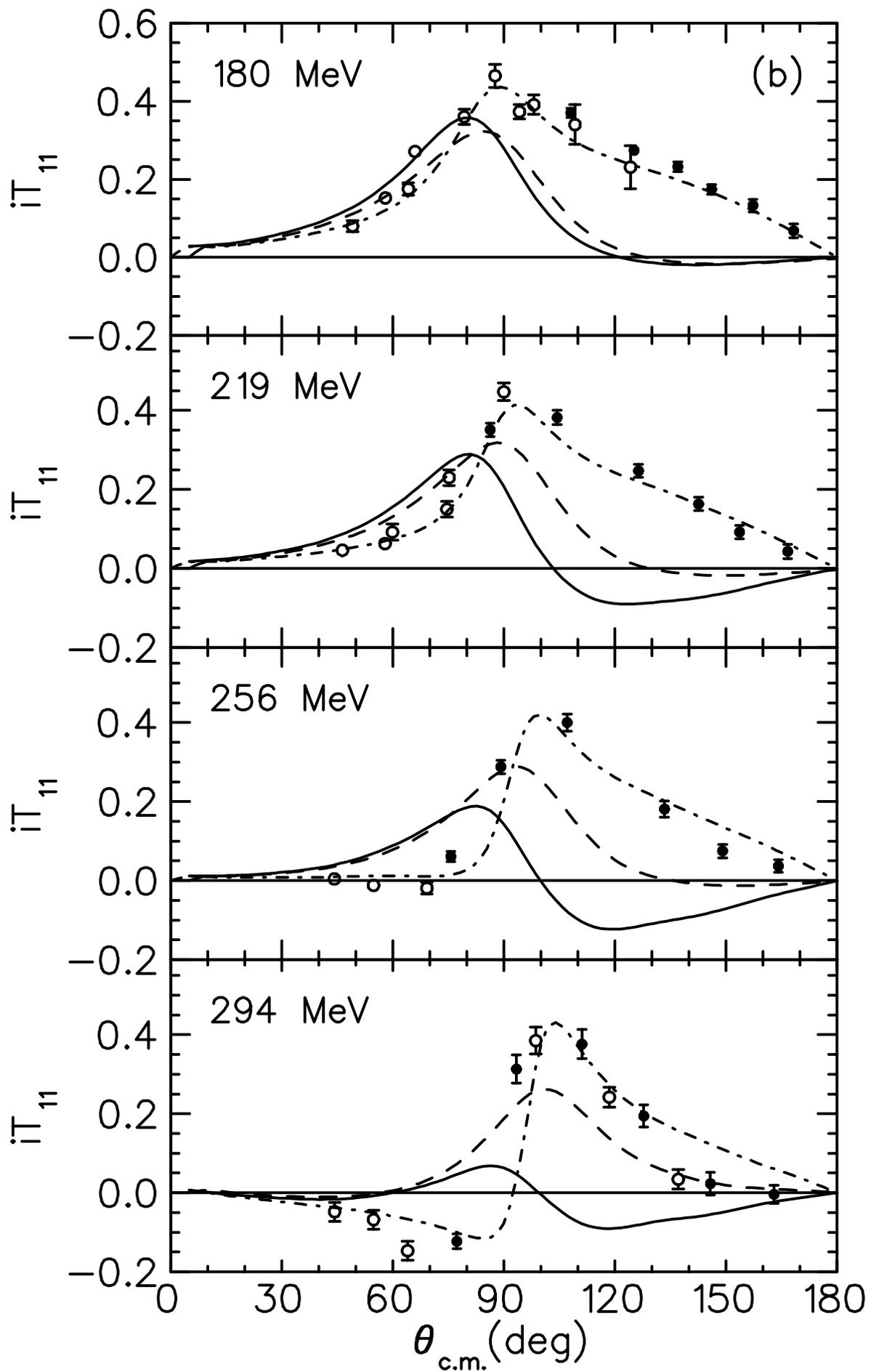
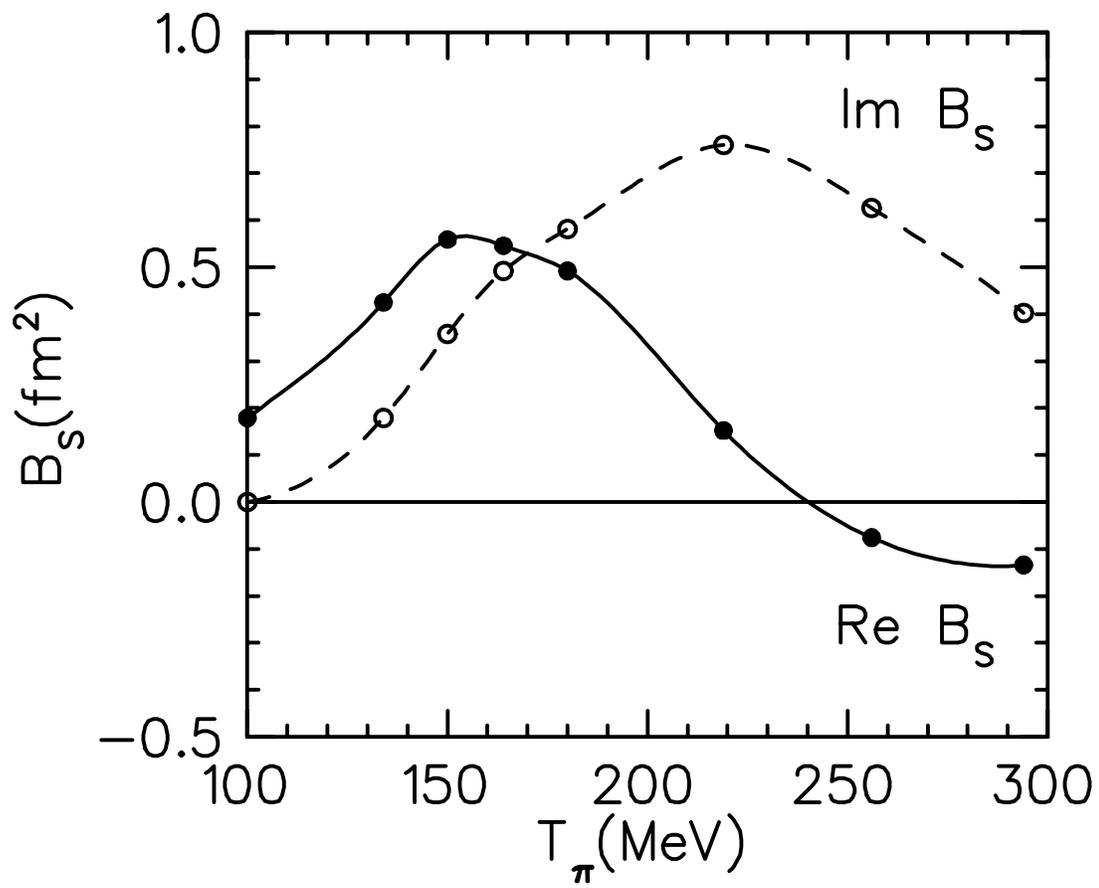
\end{document}